\documentclass[twocolumn]{aa}

\usepackage{multirow}
\usepackage{amsmath}
\usepackage{natbib}
\usepackage{bm}
\usepackage{graphicx}

\usepackage{dcolumn}

\newcommand{\om}{\Omega_{\rm M}}

\begin{document}

\title{Dark energy constraints from a space-based supernova survey}

\titlerunning{Dark energy constraints from a space-based supernova survey}
\authorrunning{P.~Astier et al.}

\author{
P.~Astier\inst{1},
J.~Guy\inst{1},
R.~Pain\inst{1},
C.~Balland\inst{1,2}
}

\institute{
Laboratoire de Physique Nucl\'eaire et des Hautes Energies, UPMC Univ. Paris 6, UPD Univ. Paris 7, CNRS IN2P3, 4 place Jussieu 75005, Paris, France
\and
Universit\'e Paris-sud, Orsay F91405, France.
}

\date{Received Mont DD, YYYY; accepted Mont DD, YYYY}

\abstract
{} {We present a forecast of dark energy constraints that could be obtained
from a large sample of distances to Type Ia supernovae detected and measured from space.} 
{We simulate the supernova events as they would be 
observed by a EUCLID-like telescope with its two imagers, assuming
those would be equipped with 4 visible and 3 near infrared swappable filters.
We account for known systematic uncertainties affecting 
the cosmological constraints, including those arising through 
the training of the supernova model used to
fit the supernovae light curves.}
{Using conservative assumptions and Planck priors, we find that 
a 18 month survey
would yield constraints on the dark energy equation of
state comparable to the cosmic shear approach in EUCLID: a variable 
two-parameter equation of state can be constrained to $\sim 0.03$ at 
$z \simeq 0.3$.
These constraints are derived from distances to about 13,000 supernovae
out to $z=1.5$, observed in two cones of 10 and 50 ${\rm deg^2}$.
These constraints do not require measuring a nearby supernova sample from the ground.}
{Provided swappable filters can be accommodated on EUCLID, distances to supernovae can be measured from space and contribute to obtain the most precise constraints on dark energy properties.}


\maketitle

\section{Introduction}
\label{sec:introduction}

About ten years ago, distances to about 50 Type Ia supernovae (SNe~Ia) enabled two teams 
\citep{Riess98b,Perlmutter99} 
to independently constrain the kinematics of the expansion of the
universe and present the first evidence for acceleration at late times
(typically later than z $\simeq$ 0.5). This acceleration was ascribed to a
mysterious component baptised dark energy. The mere cause of this late
acceleration remains unknown, and the expansion history is not
well enough constrained yet to uniquely characterise the phenomenon.
There is no known candidate to incarnate this dark energy, and working
groups have been constituted to review possible observational
strategies that could significantly increase our knowledge of it. 
Two working group reports \citep{DETF06,ESO-ESA} have identified a set of observational
approaches to constrain dark energy, and outlined generic large and
difficult projects, mainly (but non only) space-based, that could significantly
improve our knowledge of dark energy.

Four main dark energy probes were identified: the cosmic shear
correlations as a function of angle and redshift, the
measurement of the acoustic peak in the galaxy correlation function, 
the measurement of distances to Type Ia
supernovae, and cluster counts.
All these measurements have to be carried out as a function of redshift in
order to constrain dark energy.
Forecasts and merits of these methods were discussed in both reports
which stress that crossing methods is mandatory,
since the anticipated measurements face partly unknown systematic 
uncertainties. Regarding the merits of the considered probes, 
a short summary of the findings
is that cosmic shear correlations are the most promising and 
most demanding approach,
baryon acoustic oscillations (BAO) is the simplest (in terms of analysis), 
and distances to supernovae the most mature.
One strong incentive to use multiple probes, beyond redundancy,
is that General Relativity predicts a specific relation between the 
expansion history and the growth of structures that can be 
uniquely tested by comparing expansion history from supernovae and BAO,
to growth rate from cosmic shear, cluster counts and redshift distortions.
(The measurement of redshift distortions can be extracted from the same data as BAO.)

   The Hubble diagram of Type Ia supernovae remains today the best dark 
energy probe, at least 
if one focuses on results rather than forecasts. Ambitious space-based 
supernova programs have been imagined and presented right after the 
discovery of accelerated expansion\footnote{see e.g. http://snap.lbl.gov/}
that would observe O(1000) supernovae from space.
More recently, a small visible-imaging mission concept DUNE has been 
considering the observation of about 10,000 supernovae up to $z=1$ from space 
\citep{DUNE06}. This mission concept was later extended to the 
near infrared (NIR) 
and proposed to ESA under the same name \citep{DUNE08}. New space missions are 
currently being developed
in order to constrain dark energy, both in north America (the JDEM
mission, now called WFIRST) and in Europe (the EUCLID mission, which incorporates the 
second DUNE concept).  We propose here a space-based supernova survey that 
could plausibly be implemented on the EUCLID project, provided a 
filter wheel is accommodated into the visible camera. For the first DUNE
concept, suppressing the filter wheel had been considered and had a negligible
impact on the mission cost and complexity. The supernova survey forecast   
we present here relies on the experience gained in analysing the ground-based SNLS survey
\citep{Guy10, Conley10}.

We will first describe the instrument suite we simulate.
We follow in \S \ref{sec:supernova_survey} with the proposed supernova survey. 
The proposed methodology for the analysis and the associated Fisher matrix are the subjects of \S \ref{sec:methodology}. Our results are presented in \S \ref{sec:results}.
We discuss in \S \ref{sec:bias} how the key issues related to 
distance biases might be addressed. We summarise in \S
\ref{sec:discussion}.

\section{Instrument suite}
\label{sec:instrument}

In its current design, EUCLID is equipped with a
visible and a near infrared (NIR) imager \citep{EuclidYB}.
(It also embarks a slitless NIR spectrograph, and the
SN survey we discuss here does not rely on it).
We assume that both imagers
observe simultaneously the same part of the sky through a dichroic,
but if they were instead observing contiguous patches, it would marginally
affect the efficiency of the proposed surveys.

\subsection{General features}
We assume the following figures and features : 

\begin{itemize}
\item The primary mirror diameter is 1.2m, with a central occultation of 0.5m.
\item Both the visible and NIR imagers cover 0.5 ${\rm deg}^2$. The survey efficiency
driver is the field of view of the NIR imager. 
\item The pixel sizes are 0.1\arcsec\  for the visible channel and 0.3\arcsec\ for the NIR channel.
\item We assume a quantum efficiency of 0.8 for the NIR detectors. The CCD sensors for the visible channel are assumed to have a quantum efficiency peaking at
0.93 at 600 nm, and falling to 0.5 at 900 nm. The red sensitivity is probably conservative.
\item We assume that the NIR imager has a 20 electrons read-out noise and 
a 0.1  $\mathrm{el./s/pix}$ dark current. Both numbers are not particularly 
optimistic.
\item The visible imager has a 4 electrons read-out noise and a 
0.002 $\mathrm{el./s/pix}$ dark current. Both have a negligible impact on broad-band 
photometry of faint sources.
\item At variance with the current EUCLID design, we assume that 
both imagers are equipped with filters, with a filter changing device
such as a filter wheel.
\item The filters transmit at most 90\% of the light, and the optics transmission (excluding filter and CCD) 
is about 80\% at 1.2 $\mathrm{\mu m}$.
\item The image quality (IQ) is due to two components : one 
independent of wavelength equal to 0.17\arcsec (FWHM) and a 
diffraction contribution for a 1.2 m mirror.
Within this model, the FWHM IQ increases from 0.2\arcsec\  at 0.5 $\mathrm{\mu m}$ to
0.35\arcsec\  at 1.5 $\mathrm{\mu m}$. The contribution of
pixel spatial sampling will be considered later. 
\item The light background is taken from \cite{Leinert98}. We assume that
the survey takes place at 60 degrees of ecliptic latitude, which corresponds
roughly to a 20~\% increase of background light with respect to the 
ecliptic pole.
\end{itemize}

\subsection{Filter set}

\begin{figure}
\centering
\includegraphics[width=\linewidth]{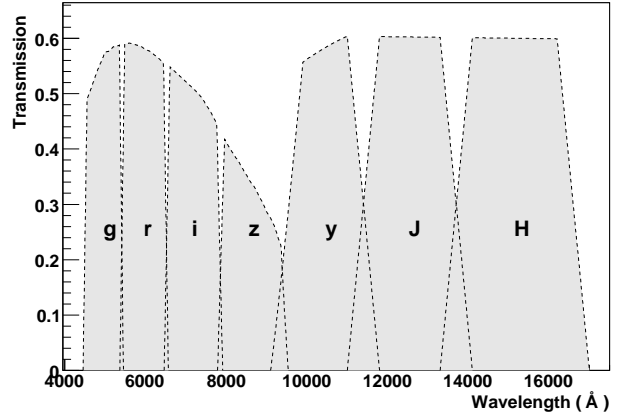}
\caption{Overall transmission of the bands of both imaging systems as they 
are simulated for this study. The EUCLID baseline set consists of 3 NIR 
filters (as assumed here) and a single ``r+i+z'' visible filter.
\label{fig:trans-filters}
}
\end{figure}

Cosmological constraints using supernovae derive from
comparing measurements across redshifts. In order to make the comparison as robust as possible,
we propose a filter set logarithmic in wavelength, with 7 bands:
4 bands in the visible channel, corresponding roughly to g,r,i and z
from the SDSS system \citep{Fukugita96}, and 3 bands on the NIR channel
roughly matching y, J and H. The NIR filters are already included in the 
EUCLID concept, but the visible filters are not. The whole filter set covers the $[450,1660]$ nm
interval, with the dichroic split at 950 nm. The overall transmission of 
the simulated bands is displayed in Fig. \ref{fig:trans-filters}. 
The sky brightness 
we assume in our simulations and the collecting power of the instrument
(provided as magnitude zero-points) can be found in Table \ref{tab:sky-zp}.

\begin{table}[h]
\begin{center}

\begin{tabular}{|c|cccc|}
\hline
band & $\bar{\lambda}$ ($\AA$) & sky (AB/$\mathrm{arcsec^2}$) & zp $\mathrm{(e^-/s)}$ & 10$\sigma$ (600s, ps)\\
\hline
g & 4965 & 23.19 & 24.27 & 25.1 \\
r & 5958 & 22.89 & 24.32 & 25.1 \\
i & 7162 & 22.68 & 24.18 & 24.9 \\
z & 8574 & 22.60 & 23.72 & 24.4 \\
y & 10476 & 22.47 & 24.32 & 24.4 \\
J & 12549 & 22.44 & 24.37 & 24.4 \\
H & 15148 & 22.31 & 24.41 & 24.4 \\
\hline

\end{tabular}
\end{center}
\caption{Characteristics of the bands considered in this work:
central wavelength,  sky brightness (in AB magnitudes/${\rm arcsec^2}$), 
zero-points (for AB magnitudes and fluxes in~${\rm e^-/s}$), 
and AB magnitude limits for point sources detected at S/N=10 in 600~s using PSF
photometry, accounting for dark current, read-out noise and pixel sampling.
Note that the sensitivities provided in \cite{EuclidYB} differ substantially because
they apply to objects measured using aperture photometry, with a higher 
sky background because the EUCLID wide survey cannot be confined to high 
ecliptic latitudes. 
\label{tab:sky-zp}}
\end{table}

\section{Characteristics of the supernova survey}
\label{sec:supernova_survey}
Supernovae are transient events and the primary distance indicator is
the amplitude of the light curve(s), that we assume to be measured via
broad-band imaging. In order to properly sample supernova light curves, one
needs to repeatedly image the same pointings, in as many bands as
possible. We require that the rest frame wavelength region from 4000 to
6800 \AA\  is measured for all events, and sampled by (at least) three
filters: this roughly corresponds to measuring the B,V and R
(rest frame) standard bands. Current distance estimators only require
two bands (they can of course accommodate more bands). Requiring
three bands will either allow one to use more elaborate distance estimators, 
or to use the built-in redundancy to compare derived distances. Requiring that the same rest frame wavelengths are measured
at all redshifts make the comparison of supernovae brightnesses as
independent as possible from a supernova model. Although we require 
three bands to measure a distance, the observing strategy proposed below
provides us with more than three bands at all redshifts. 
The duration of the survey is arbitrarily chosen as 1.5 year,
i.e. about one third of a five-year mission. This range of
survey time provides O($\mathrm{10^4}$) events, i.e. an order of 
magnitude over current
cosmological samples (e.g. \citealt{Amanullah10}). 
Alterations of the survey duration will be considered in \S \ref{sec:variations}.

\subsection{Photometric precision requirements}
\label{sec:photometric_precision_requirements}
The depth of the observations is driven by the maximum target 
redshift, and considering three requirements:
(1) the measurements should be accurate
enough so that the distance estimate has a measurement uncertainty
that remains significantly smaller than the observed scatter
of the Hubble diagram (namely around 0.15 mag for modern surveys, 
see e.g. \citealt{Guy07,Kessler09,Guy10,Conley10}); 
(2) the same supernova parameter
space should be observable at all redshifts in order to avoid the 
shortcomings of Malmquist bias corrections;
(3) the amplitude of the three bands 
roughly matching rest frame B,V,R bands should be measured 
at an accuracy better or comparable to the measured ``colour smearing'',
namely about 0.025 mag, see \cite{Guy07,Guy10}.
Colour smearing\footnote{The expression was introduced in \cite{Kessler09}.} refers
to the spread of colour-colour relations of SNe~Ia: it is the scatter of 
single-band light curve amplitudes one has to add to measurement 
uncertainties in order to properly describe the observed spread 
of colour-colour relations of SNe~Ia.
One can for example model the rest frame V-R colour from B-V. The scatter of this 
colour-colour relation is properly described
by assuming that the B,V and R peak magnitudes scatter 
independently by $\sigma_c \simeq$~0.025~mag around a two-parameter model.
One can estimate the colour smearing
as a function of wavelength (\citealt{Guy10}), and find that the B,V,R region
is less scattered than bluer bands.  The colour smearing contributes to
the Hubble diagram scatter, but does not account for the entirety of the 
measured $\sim$0.15 mag r.m.s.

   In practice, the requirement (2) about Malmquist bias
indicates that events one magnitude fainter than the average
at the highest redshift should be easily detected. We find that 
requirement (3) is the most demanding, and the survey setup we 
present later fulfils these three requirements. 

\subsection{Redshifts}
Supernova redshifts are obviously needed to assemble a Hubble
diagram.  Obtaining SNe~Ia spectra from the ground is technically
feasible up to $z\sim 1$. At higher redshifts, the required exposure
times increase very rapidly with redshift because the SNe~Ia 
flux decreases rapidly in the near UV (bluer  than $\sim$ 3600\AA),
and the atmosphere glow increases rapidly towards the red.
A large spectroscopic followup of supernovae at $z>1$
is hence out of reach of current ground-based instruments, and 
even if adaptive optics, OH suppression and larger telescopes
will certainly help very significantly, massive
statistics will likely remain out of reach during at least the next decade. 

Since acquiring a spectrum for each supernova was a core goal of the 
proposed SNAP concept\footnote{http://snap.lbl.gov}, its baseline design 
included a high-throughput low-resolution spectrograph.
In this concept, the supernova spectroscopy requirements
drove the mirror size and limited the statistics to O(2000) supernovae,
because of the long exposures required for
spectroscopy of distant supernovae. 

We believe that obtaining supernova spectra one at a time with a small field 
instrument is not a realistic goal for O($\mathrm{10^4}$) events, 
especially if aiming
at redshifts beyond unity. Wide field space-based
slitless spectroscopy may be considered, but the S/N ratio of slitless
spectra is naturally poor, because every pixel integrates
the sky background spectrum in the whole bandwidth: spectra
at a modest magnitude $H \sim 24$ seem out of reach of EUCLID slit-less spectrograph.
Space-based spectroscopy with
synthetic slits (as originally proposed for the SPACE project, 
later merged with DUNE into EUCLID) is
significantly more sensitive, and for a supernova program on a joint
imaging-spectroscopy-with-slits mission (such as the original EUCLID concept),
one should obviously consider obtaining supernovae spectra in parallel
with repeated imaging.

So, for O($\mathrm{10^4}$) supernovae, we should consider the case where supernova
spectra cannot be acquired for all events. Since
the Hubble diagram obviously requires redshifts, we consider 
the following alternatives:
\begin{itemize}

\item Acquiring host galaxy spectra ``after the fact'' using
wide field spectroscopy. The target density would be below
500 deg$^{-2}$ and hence perfectly suited to the multi-fiber
spectrography being considered today (WFMOS and BigBoss are
typical examples). For the redshift range and area we will consider below,
this would typically require a few hundred hours of integration on 
a 4-m class telescope, which seems acceptable, thanks to the multiplex factor.
\item Relying on photometric redshifts of host galaxies,
using both visible and NIR bands. 
This would cause a loss of accuracy of cosmological parameters. 
However, one would still have to collect a sample of spectroscopic redshifts
of faint galaxies in order to train the photometric redshift
algorithms.

\item Relying on photometric redshifts of supernova. These are now known
to be more accurate than photometric redshifts of host galaxies
\citep{Palanque10,Kessler10}, thanks to the homogeneity of the
events. However, this approach weakens the identification step since
the redshift will minimise the difference between measurements and
expectations for a Type Ia supernova. It also introduces correlated
uncertainties between distance and redshift which would require a
careful study.
\end{itemize}

We assume in what follows that supernovae have high quality redshifts,
i.e. that spectroscopic redshifts of host galaxies are acquired.
From here on, we will allow for a conservative 20 \% loss in supernovae 
statistics due to this process.

\subsection{Why space?}
Two aspects favour a space based supernova survey: first, ground-based
surveys suffer from variations of atmospheric transparency and 
image quality across observing epochs. This causes photometric uncertainty 
floors which can be overcome from space. Second, photometry
of faint sources in the NIR from the ground is notoriously
difficult and requires extreme exposure times on 8-meter class 
telescopes, because of the bright atmospheric glow. 
The ``1 $\mathrm{\mu m}$ barrier'' practically limits 
supernova ground-based surveys to $z=1$ or significantly below, if one imposes
that supernovae at all redshifts are measured in the same rest frame wavelength range.
Space offers NIR photometry with sensitivities similar to the
visible range, and hence opens supernova surveys to the $z>1$ range.
Accessing NIR bands also improves supernova surveys in two respects:
it allows one to measure objects on a large wavelength lever arm,
which is mandatory to characterise their colour variations.
It also allows one to measure distances to supernovae using redder 
bands than allowed from the ground (we will consider  
rest frame I band in \S \ref{sec:I-band}), which limits the effects of colour variations among
the sample. Finally, accumulating deep NIR photometry
of galaxies improves galaxy typing and photometric redshifts (over visible-only measurements);
the latter may be useful even for supernova cosmology if obtaining 
spectroscopic redshifts of the supernova hosts turns out to
be not practical.

\subsection{Supernova surveys cadence and coverage}
\label{sec:cadence_and_coverage}

The imaging survey should be run in ``rolling search mode'' where 
a given patch of the sky is observed repeatedly, in order to discover
and measure variable objects. The required wavelength range to cover
is bound to the redshift range aimed at: in order to derive
distances as independent as possible of any supernova model, the same 
rest frame
spectral region should be used to derive distances at all redshifts.

Monitoring a single cone at the depth required at the survey highest
redshift delivers a supernova redshift distribution where moderate and
low redshifts are essentially missing. We hence propose a two-cone
approach: a deep survey of 10 ${\rm deg^2}$ up to $z=1.55$, and a wide
survey of 50 ${\rm deg^2}$ up to $z=1.05$. We consider different 
area ratio in \S \ref{sec:variations}. The deep survey typically
requires integrations 4 times longer than the wide. 
Note that both surveys are on purpose volume limited in order to avoid
the shortcomings of selection bias corrections.

We simulate a survey duration of 1.5 year of calendar time, with 
the wide and deep survey observations interleaved. The footprint 
of the deep survey is assumed to be outside the wide, so that its low redshift sample
(where statistics is precious) adds up to that of the wide.
We draw the simulated samples from a measured SNe~Ia
volumic rate as a function of redshift (\cite{Ripoche08}, 
see also \citealt{Perret10}), which may
be parametrised as 
$$ 
R(z) = 1.53\ \ 10^{-4} \left [ (1+z)/1.5 \right ]^{2.14} h_{70}^3\  \mathrm{Mpc^{-3}\  yr^{-1}} 
$$ 
where years should be understood in the rest frame. 
Since these measurements stop around $z=1$ and rates at larger
redshifts are highly uncertain, we assume that the volumic
rate remains constant above $z=1$ (to $z=1.5$). The rates proposed in
\cite{Mannucci07} (accounting for events ``lost to extinction'') yield
a statistic (to $z=1.5$) $\sim$ 25 \% larger than what we simulate,
with a similar redshift distribution. 
Our simulation accounts for ``side effects'' by 
requiring that all epochs corresponding to rest frame phases from -15 to +30 
days from maximum light are measured. As a consequence the event statistics increases
if the survey monitors smaller areas over a longer period, with a 
constant total observing time.  
The accuracies we discuss later ignore $20 \%$
of the events, in order to allow for losses in the measurement process
(failures to obtain redshifts for example). Accounting for these losses,
the deep and wide surveys deliver
about 4000 and 9000 events respectively. The simulated statistics
are provided as a function of redshift in Table \ref{tab:nsn}.
Events at higher redshifts can be detected but they fail
the quality cuts for deriving distances.
\newcommand{\nodata}{\multicolumn{1}{c}{-}}
\begin{table}[h]
\begin{center}
\begin{tabular}{|c c| r r |}
\hline
$z_\mathrm{min}$ & $z_\mathrm{max}$ & wide & deep  \\
\hline
0.15 & 0.25 & 120 & 24  \\
0.25 & 0.35 & 263 & 52  \\
0.35 & 0.45 & 454 & 90  \\
0.45 & 0.55 & 685 & 137  \\
0.55 & 0.65 & 945 & 189  \\
0.65 & 0.75 & 1226 & 245  \\
0.75 & 0.85 & 1520 & 304  \\
0.85 & 0.95 & 1819 & 363  \\
0.95 & 1.05 & 2090 & 418  \\
1.05 & 1.15 & \nodata & 435  \\
1.15 & 1.25 & \nodata & 440  \\
1.25 & 1.35 & \nodata & 442  \\
1.35 & 1.45 & \nodata & 439  \\
1.45 & 1.55 & \nodata & 434  \\
\hline  
\multicolumn{2}{|c|}{total} & 9122 & 4012  \\
\hline
\end{tabular}
\end{center}
\caption{Expected number of SNe~Ia events in 1.5 year
for the wide and deep surveys respectively (accounting for
a spectroscopic efficiency of 0.8). Events beyond a redshift
of 1.05 in the wide survey are ignored because the quality of 
their measurements is below our requirements. \label{tab:nsn}}
\end{table}

We settle for a cadence of 5 observer days, but the photometric
accuracies that matter to derive distances do not depend on this choice at first order, since
those mainly depend on the overall integrated light. However, 
a significantly coarser sampling might compromise photometric identification.
The integration times per filter at each epoch are provided in Table~\ref{tab:exp-times}
 and were chosen in order to provide a measurement
of light curve amplitudes to a precision of 2.5 \% r.m.s on average at the
highest redshifts of each survey.
\begin{table}[h]
\begin{center}
\begin{tabular}{|c|c c|}
\hline
Filter & wide & deep \\
\hline 
g     & 400  &  1600 \\
r     & 400  &  1600 \\
i     & 500  &  1600 \\
z     & 700  &  2400 \\
\hline 
y     & 650  &  1800 \\
J     & 650  &  2400 \\
H     & 650  &  3000 \\
\hline
\end{tabular}
\end{center}
\caption{Integration times per visit (in seconds) proposed for the wide and 
deep supernova surveys with a 5 days cadence. The wide and deep surveys have
to cover respectively 100 and 20 pointings. The CCD type we simulate
has a modest quantum efficiency in the z band, which drives exposure
times up in this band.
\label{tab:exp-times}}
\end{table}

In 5 days of calendar time, the total exposure times amount to 56 and 40 hours for the wide
and deep surveys respectively, which corresponds to 80~\% of the 120
hours available.

Both surveys are conducted in 7 bands, covering g to H 
($450<\lambda<1700\ \mathrm{nm}$). One may
argue that the bluest bands of the deep survey and the reddest bands
of the wide survey are not strictly needed to measure supernovae distances, but since
visible and NIR observations are assumed to happen in parallel,
dropping blue visible bands for the deep survey, or red NIR bands for the wide
survey does not save any observing time.

Note that we have not discussed the collection of a nearby sample 
(typically O(1000) events at $0.03<z<0.1$). As a baseline, 
and at variance with the SNAP project, 
we stick to a self-contained imaging survey, in order to realistically
limit cross-calibration and detection bias issues. 

\subsection{Identification of SNe~Ia and contamination}
SNe~Ia exhibit reproducible rest frame colours ($\sigma(B-V) \simeq 0.1$, see e.g. Fig. 8 of \citealt{Astier06}), 
and even more reproducible colour relations. For example, the rest frame 
U-band amplitude can be predicted to better than $\sim$ 0.04 from 
B- and V-bands \citep{Astier06}. SNe~Ia not only occupy narrow subspaces of 
multicolour spaces, but also exhibit  very reproducible light curve shapes 
that permit to discriminate against most of the core-collapse events \citep{Poznanski02,JohnsonCrotts06,Rodney09}.
Studies of the photometric selection of SNe~Ia have been 
conducted on the SNLS data and their preliminary conclusions are
encouraging \citep{RipochePHD,BazinPHD}.  These studies rely on host galaxy 
photometric redshift and most of their identification failures are due 
to wrong assumed redshifts. The supernova survey we are considering here
would be in a more favourable situation than these studies: it measures 
7 bands (whilst SNLS has at most 4), and we assume that host galaxy 
spectroscopic redshifts will be
available (whilst SNLS studies used host galaxy photometric redshifts).

In order to estimate the contamination by core-collapse supernova
in a sample selected in colour-colour subspaces, we would need
a large enough sample of multi-band measurements of such events.
These should soon be available, thanks at least to
the SDSS supernova survey, the Lick Observatory Supernova Search, 
and the Palomar Transient Factory, but they are not available yet.
Therefore, in order to bound the impact of core-collapse contamination 
on the SNe~Ia distance-redshift relation, we 
resort to studying how clipping around the Hubble line rejects
other supernova types, following \cite{Conley10}.

Core collapse supernovae are classified in Type Ib and Ic, and Type II.
Type Ib and Ic events are often merged into a ``Ibc'' type
(see e.g. \citealt{Richardson02, Li10II}). 
Type II events are about 3 times more frequent than Ibc (\citealt{Li10II} Fig. 9),
but two thirds of those are Type II-plateau (II-p) which are easily identified 
from their very flat light curves. Other Type II events represent 
about the same rate as Ibc, but their light curves rise in a few days, whereas
SNe~Ia rise in more than 15 days. So, Type II events might add a small 
contribution to Ibc interlopers, and we will now concentrate on evaluating the
impact of a Ibc contamination in the Hubble diagram.

We model the Ibc population absolute magnitude distribution as a Gaussian
offset by $\Delta_{bc}$ from the Ia population with r.m.s $\sigma_{bc}$.
We expect the rate of Ibc to be proportional to the star formation rate, that
we take from \cite{Hopkins06}, and the amount of Ibc events follows from $f_{bc}(z=0)$,
the ratio of the Ibc rate to the Ia rate at $z=0$.
The adopted Ia rate was presented in \S \ref{sec:cadence_and_coverage}.

  We simulate a mix of Ia and Ibc events, fit the Hubble diagram with a 
sixth degree polynomial, and iteratively clip events beyond 3 $\sigma$ from
the fit and refit, until no event is clipped.
We simulate Ia events with a Gaussian scatter 0.15 mag around the Hubble line.
For Ibc events, it is unclear how applying to them
the brighter-slower and brighter-bluer corrections for SNe Ia will affect
the absolute magnitude distribution. We might guess that some part
of their brightness scatter is due to extinction in their host galaxy,
and that brighter-bluer corrections would narrow their distance modulus distribution.
We will consider below two estimates of the Ibc brightness scatter,
one as observed and one significantly lower. Note that \cite{Richardson02}
estimate intrinsic magnitude scatter corrected for host galaxy extinction
which differ little from raw estimates. 
The contamination depends on the bright end
of the Ibc luminosity function which is not well constrained: 
\cite{Richardson02} propose a distribution extending well beyond the
Ia average brightness, while \cite{Li10II} brightest Ibc event is 0.5 mag fainter
than the average Ia. Both results are however
compatible at the $\sim$10 \% CL, given the modest statistics 
involved\footnote{Assuming the disagreement is real, it might be due to
the fact that the \cite{Li10II} sample comes from a search targeting 
nearby galaxies and could be missing events preferentially occurring in 
dwarf galaxies.}. We propose three scenarios for the Ibc population, each 
with two values for $\sigma_{bc}$, detailed in Table
\ref{tab:contamination}:
\begin{itemize}
\item (R02) the Ibc population in \cite{Richardson02} amounts to
about 16 \% of the SNe~Ia, is fainter by 1.4 mag than SNe Ias 
and has a r.m.s.  scatter of 1.4 mag.
\item (R02 bright) \cite{Richardson02} see a mild indication of a bright
component of Ibc, representing about 5 \% of the SNe~Ia.
\item (L10) \cite{Li10II} have a much more complete survey and find a volumetric 
rate of Ibc which is about 80 \% of the Ia rate. Their average Ibc is fainter
by 2.4 mag than their average Ia with an r.m.s of 1.2 mag. 
as we just noted, the magnitude distribution of the measured sample does
not contain Ibc events under the Ia peak.
\end{itemize}
Contamination affects cosmology by biasing the average distance modulus.
However, a redshift independent bias has no effect
on cosmology, and we report in Table \ref{tab:contamination} the slope of the 
bias as a function of redshift $d\,\delta\mu/dz$, which fairly describes 
most of the effect. 

\begin{table}[h]
\begin{center}

\begin{tabular}{|l|ccccc|}
\hline
Scenario                   & $\Delta_{bc}  $ &   $f_{bc}$  &  $\sigma_{bc}$ & $d\,\delta\mu/dz$ & $N_{bc}/N_{Ia}$  \\ 
\hline
\multirow{2}{*}{R02}       &\multirow{2}{*}{1.4}  & \multirow{2}{*}{0.16} & 1.4 & 0.0014 & 5.5\% \\
                           &                      &                       & 1.0 & 0.0034 & 4.9\% \\
\multirow{2}{*}{R02 bright}&\multirow{2}{*}{-0.8} & \multirow{2}{*}{0.05} & 0.5 & -0.0047 & 2.6\% \\
                           &                      &                       & 0.33& -0.0047 & 1.6\% \\
\multirow{2}{*}{L10}       & \multirow{2}{*}{2.4} & \multirow{2}{*}{0.8}  & 1.2 & 0.0086 & 8.4\% \\
                           &                      &                       & 1.0 & 0.0058 & 4.3\% \\
\hline 
\end{tabular}
\end{center}
\caption{
Various hypotheses about the average magnitude shift of Ibc's with
respect to SNe~Ia, the r.m.s of their magnitude distribution, the
amount of Ibc at $z=0$ in units of the Ia rate, the resulting slope
bias of the Hubble diagram $d\,\delta\mu/dz$, and the average fraction of
contaminants. The impact on cosmology is small under all these hypotheses.
\label{tab:contamination}}
\end{table}

We find that under the three scenarios, the effect on cosmology is small.  Even in
our ``L10'' scenario, where the contamination is suspiciously large
compared to the fraction of Ibc identified by high redshift SNe~Ia
surveys, the effect on cosmology remains below the systematic
uncertainty $\sigma(e_M)=0.01$ (defined later in
Eq. \ref{eq:M_depend_z}) that we will consider as a baseline. The
clipping process has a negligible impact on the SNe~Ia statistics. A
rough estimate of the size of the effect we find can be readily
computed, for e.g. the first line of the table: the average offset of
Ibc within the Ia $\pm 3\sigma$ window is 0.047, the fraction of the
Ibc population within the same window amounts to $\sim$20\%, the Ibc
to Ia number ratio at $z=0$ is assumed to be 0.16, and it increases with
redshift as $\sim(1+z)$.  The evolution of the Ibc induced distance
bias reads $0.047*0.2*0.16 = 0.0015$. 

One might argue that a Gaussian
distribution is inadequate to describe the tails of the Ibc
distribution, but more populated tails also exhibit shallower slopes,
and both alterations have a tendency to cancel each other. Note that
our estimates ignore the potential rejection from light curve shapes
and colours of Ibc events, which is certainly a conservative
assumption.
 
Our results might look at odds with other attempts. \cite{Homeier05}
follows a similar approach and finds sizable effects on cosmology, and
we attribute the difference to the absence of clipping.  More
recently, \cite{SnIdChallenge} proposed a supernova photometric
identification challenge. The provided simulated sample contains a
large fraction of events for which the phase coverage and the signal
to noise ratio are inadequate for a distance measurement, would they
be genuine Type Ia. In our simulation, such events are excluded a
priori and do not count as identification failures. We also benefit in
our simulation from assuming spectroscopic redshifts as this reduces
the Hubble diagram scatter. Regarding the identification using
colours, Type Ia events were generated in \cite{SnIdChallenge} with a
colour smearing of 0.1 mag, which we now know to be about 4 times too
large, and leads to overestimating mis-identifications.

We eventually ignore the efficiency loss due to photometric selection, 
and integrate the effects of possible contaminations into a drift of
absolute magnitude (see Section \S \ref{sec:distance_estimation_parameters}
and Eq. \ref{eq:M_depend_z}).
Note that if spectroscopic redshifts of host 
galaxies are acquired and the wrong host is assigned to a supernova, 
this event will likely fail the photometric typing cuts and hence will 
not pollute the Hubble diagram.

\subsection{Rest frame I-band Hubble diagram}
\label{sec:I-band}
The wide survey provides the possibility of constructing a rest frame
I-band Hubble diagram out to $z\simeq 0.9$, with a statistics of about 7000
events. Encouraging cosmological results
were recently obtained from rest frame I-band measurements from the ground
\citep{Freedman09}. I-band distances have a smaller contribution of colour to the
distance estimate than B-band distances, and are hence significantly
more robust to systematics related to colour modelling and calibration.
They also exhibit a smaller scatter. 
Measuring distances to the {\it same} events using independent
measurements constitutes a very appealing test of the methods and
possible biases. It seems unlikely that collecting full rest frame
I-band light curves from the ground for a O(1000) events sample reaching
$z=0.9$ becomes feasible in the next decade. In what follows, we 
conservatively did not include the cosmological constraints that could 
be obtained from this second built-in supernova survey.

\section{Methodology}
\label{sec:methodology}

\subsection{Point source photometry}
In deriving the photometric accuracy of a measurement, we assume that the 
photometry of supernovae is carried-out using
PSF photometry\footnote{We assume that the object position is perfectly known:
first it does not improve the flux variance, and second, we are studying 
here a  supernova survey where the object is measured at the same position 
in an image series.}. 
The pixel sampling is accounted for, and we checked that
the position of the source centre within a pixel does not change significantly
the signal to noise ratio. For most of our
supernovae measurements, the contribution of the shot noise from 
the source itself is not negligible. 

\subsection{Light curve fitter training}
We use the SALT2 model \citep{Guy07} to generate light curves. In
this framework, light curves are described by 4 parameters: a date of
maximum light (in B-band), an overall amplitude of the light curve
$X_0$, an $X_1$ parameter closely related to stretch factor or decline
rate, and a colour at B maximum. Since light curve shapes are not
derived (yet) from explosion models, the light curve models rely on a
training sample, and then reflect the quality of this training
sample. Since we should aim at obtaining a supernova sample significantly
larger and of better quality than before, the sample itself should be
used to train the light curve fitter. This self-training procedure is
only possible for light curve fitters which do not provide distance
estimates, such as SALT and SALT2 \citep{Guy05,Guy07}, SIFTO
\citep{Conley08}. Those do no assume any relation between redshift and 
brightness and only predict light curve
shapes in observer bands and flux ratios between bands or epochs. They
leave to a later stage the event parameter combination that will be
used as a distance estimator.  Supernova model training causes
specific uncertainties: model noise due to the finite training sample
and the calibration uncertainties of the latter, and a specific
distance uncertainty pattern due to self-training.

SNe~Ia exhibit a variety of rest frame colours, but different colours
of the same event seem closely related. To account for correlated 
colour variability, one can take guidance from light extinction (by e.g. dust)
and model the observed magnitude (for example at time of maximum B light) as:
\begin{equation}
m_{obs}(\lambda) = m_0 + m_1 (\lambda) + { c\ C_l(\lambda)}
\label{eq:colour_law}
\end{equation}
$m_0$ and $c$ vary from event to event, whilst the ``naked'' flux $m_1$ and 
the colour law $C_l$ constitute the model (supposedly for all events). 
We can chose the scaling
of $C_l$ so that $c = (B-V)_{max} + constant$. $C_l$ (called the 
``colour law'' in the SALT2 parlance) determines the colour relations, 
e.g. the slopes measured in rest frame colour-colour planes. 
When a similar model is developed for extinction by dust, 
the equivalent of the $C_l$ function is determined from data. 
For example 
\cite{Cardelli89} propose a one-parameter family of extinction laws
commonly called ``Cardelli laws''.
For a supernova model, $C_l$ can also be extracted from data,
however up to an arbitrary constant which does not affect
colour-colour positions\footnote{The $C_l(\lambda)$ family 
reported in \cite{Cardelli89} for extinction by dust are also obtained 
from  colour-colour slopes. The arbitrary constant can however be
determined by requiring that $C_l(\lambda)$ is 0 for $\lambda$ going
to infinity. Observations extending to $\sim \mathrm{5\ \mu m}$ allow
the authors  to carry out the extrapolation. This is not possible 
in the supernova  framework and $C_l(\lambda)$ is extracted from 
colour-colour relations  up to an additive constant. SALT2 chooses 
$C_l(\lambda_B) = 0$.}. \cite{Guy05,Guy07} derive
$C_l$ for SNe~Ia and find it significantly different from Cardelli laws. Similarly,
the SNe rest frame colour relations (e.g. U-B vs B-V) found 
in \cite{Conley08} are different from the ones expected from Cardelli laws.
  
   The Cardelli laws not only predict colour-colour slopes,
but they also relate colour variations to total extinction.
Extinction causes a brighter-bluer relation, characterised
by $R_B$=4.1 on average for the Milky Way dust. The brighter-bluer relation
can again be measured for supernovae (by minimising the 
Hubble diagram residuals), and one finds a value around 3.2 \citep{Guy10}, 
significantly
smaller than 4.1. We will come back to this point in \S \ref{sec:distance_estimation_parameters}.

Since both SNe colour relations and total to selective extinction point
away from extinction laws for Milky Way dust, we should make 
provision in the survey
design for studying both aspects using supernovae, and in particular
be in a position to measure both colour relations and the brighter-bluer
correlation without any assumption nor prior. It is essential not to 
restrict the possibilities of SN colour variations to those described
by the Cardelli laws.
Even if dust extinction parametrisations adequately described 
supernova colour relations, the known variations of dust properties 
would still require precise supernova measurements to define 
which dust is in action: the
fact that ``regular'' dust is unlikely to be the first cause of
supernova colour variations does not really impact on the
requirements of a supernova survey.
The causes of SNe~Ia colour variability are unclear, and
collecting as many colours as possible for each SN event
will likely improve our understanding of these matters.

As noted above, the data scatters around 
these empirical modelling of SNe~Ia colour relations beyond
measurement uncertainties (\citealt{Guy07,Guy10}). 
We referred to this extra noise 
as colour smearing and used a value of $\sigma_c=0.025$~mag
to set the depth of the observations. This finite colour smearing
might indicate a fundamental limitation of extinction-inspired models
to describe SNe~Ia colour variability.

\subsection{Fisher Matrix}

We want to forecast uncertainties of cosmological parameters
expected from various survey scenarios, and various hypotheses
regarding sources and size of systematic uncertainties. The Fisher 
matrix framework is sufficient for our purpose. 

Sources of uncertainties can be regarded
either as extra noise or as extra parameters, and the choice is
only a matter of convenience, as discussed on a practical example
in Appendix \ref{app:succ-simul}. We settle for adding parameters 
corresponding to uncertainty sources, and consider 6 parameter sets:
\begin{enumerate}
\item the cosmological parameters.
\item the parameters of the SN events themselves.
\item parameters describing the (rest frame) colours of an average supernova.
\item parameters describing the colour law, i.e. how these colours change from event to event.
\item parameters describing the brighter-slower and brighter-bluer
relations, and the intrinsic brightness of a supernova.
\item the photometric zero-points of the light curve measurements, 
i.e. how instrumental fluxes are converted to physical fluxes.
\end{enumerate}

We use least squares estimators. Usually, supernova cosmology fits are
carried out in two successive steps: the measured light curves are
first fit to yield the light curve parameters, and the cosmology is
then fitted using these light curve parameters (and their
uncertainties). We use here a mathematically equivalent approach where
a single simultaneous fit considers all parameters at once.  The
Appendix \ref{app:succ-simul} discusses why these two approaches are
strictly equivalent. We settle here for the simultaneous fit because
the same events are assumed to be used both to train the light curve
fitter and to measure cosmology. In particular, photometric
calibration uncertainties affect cosmology both directly and through
the light curve fitter training. Within a simultaneous fit,
propagating the uncertainties (with correlations between events) does
not require any particular care. One obvious drawback of the
simultaneous fit is that considering the supernova event parameters on
the same footing as all the other ones considerably increases the size
of the least-squares problem and uncomfortably lengthens the required
matrix inversion (as experienced in \citealt{Kim06}). In Appendix
\ref{app:local-global}, we describe how we take advantage of the
specific structure of our least squares problem to rephrase the linear
algebra using only small-sized matrices.

We now enter into the details of the chosen parametrisations.

\subsubsection{Dark energy parameters}
We follow the commonly used equation of state (EoS) effective parametrisation due
to \cite{ChevallierPolarski01}: $w(z) = w_0 + w_a\ z/(1+z)$. 
\subsubsection{Event parameters}
We use the SALT2 parameters: an overall amplitude,
a parameter very close to stretch-factor, 
a date of maximum (B-band) light, and a rest frame colour at maximum.
\subsubsection{Average magnitude of a supernova and colour law} 
Following the SALT2 model and Eq. \ref{eq:colour_law}, we model the peak magnitudes 
of supernovae as:
\begin{equation}
m(\lambda) = m^*_B+m_1(\lambda) + c C_l(\lambda)
\end{equation}
where $\lambda$ refers to a rest frame wavelength. The two functions 
$m_1$ and $C_l$ describe respectively the magnitude of an average
supernova as a function of rest frame wavelength and the colour law.
We chose, $m_1(\lambda_B) = 0$, $C_l(\lambda_B) = 0$
and $C_l(\lambda_V) = -1$, so that $m^*_B$ represents the actual magnitude
in rest frame B band, and $c = B-V + \mathrm{constant}$ is the colour of the event.
Determining $m_1$ and $C_l$ from the data is part of the light curve fitter
training. Rather than modeling the functions themselves, we model offsets
to the same functions extracted from the SALT2 model, by defining:
\begin{align}
m_1(\lambda) &\equiv m_{1,SALT2}(\lambda) + p_1(\lambda) \nonumber \\
C_l(\lambda) &\equiv C_{l,SALT2}(\lambda) + p_2(\lambda)
\label{eq:polys}
\end{align}
where $p_1$ and $p_2$ are polynomials with $N_1$ and $N_2$ parameters 
to be fitted respectively. The role of these two parametrised 
functions is to correct
broadband inaccuracies of an otherwise correct ``narrow band'' modelling 
of supernovae and their colour variations. These inaccuracies are typically
expected from photometric calibration errors of the training sample,
and do not require sharp corrections. We settle for 10 parameters 
for each polynomial, which covers spectral resolutions coarser than $\sim$ 20.
Our results do not degrade significantly with larger values, up to $\sim$ 20
parameters, where the Fisher matrix becomes numerically singular, because the 
broadband data does not contain enough information at these higher 
spectral resolutions.

   These two polynomials can be defined either as affecting the amplitude
of a light curve (at the central wavelength of the considered filter)
or as multiplicative factors affecting the supernova flux before integration
in the passband. We eventually settle for the second approach because it is
more realistic.

\subsubsection{Distance estimator parameters} 
\label{sec:distance_estimation_parameters}
The simplest way to model the 
brighter-slower and brighter-bluer correlations 
is via linear coefficients in the distance modulus
(see e.g. \citealt{Tripp98,Astier06}):
\begin{equation}
\mu = m^*_B + \alpha X_1 - \beta c - \mathcal{M}
\label{eq:m-alpha-beta}
\end{equation}
where $X_1$ may either be the actual $X_1$ parameter from SALT2 or any
other empirical parameter describing light curve width. $\mathcal{M}$ is the
intrinsic B-band magnitude of a supernova of null $X_1$ and c. Note
that a rest frame band different from $B$ can be chosen, 
as well as a different colour, as in \cite{Freedman09}. All
supernova cosmology works regard the brighter-slower relation
($\alpha$) as empirical and explicitly or implicitly fit for
it. Putting the $\beta$ parameter on an equal footing is much less
common because the assumption that colour variations among supernovae are
caused by dust extinction only readily provides a value (or a range of
values) for $\beta$ together with a prediction for the $C_l$ function
above. Many supernova works assume that the brighter-bluer
relation of SNe, parametrised by our $\beta$ parameter, is due
to redenning by dust similar to Milky Way dust. Although, one should
indeed expect that some amount of dust extinction is at play, 
other sources might dominate colour variations among events. 
Fitting for $\beta$ constitutes a more general approach than 
assuming its value, especially since the found values depart
from Milky Way dust redenning (e.g. \citealt{Tripp98} and references therein,
\citealt{Astier06,Freedman09,Kowalski08,Amanullah10}). Assuming that colour variations
of supernovae are due to Milky Way like dust not only increases the 
scatter of distances but also may cause artefacts such as evidence 
for a Hubble bubble \citep{Conley07}. In
\cite{FOM-SWG09}, the difference between the expectation
that the brighter-bluer relation of SNe follows redenning by Milky Way dust
 and evidence from measurements that it is not the case 
is described as a systematic uncertainty. We will not follow this route:
since distances
depend on this measurable $\beta$ parameter (or its equivalent in a more
complex distant estimator), it has 
to be measured, whatever the source of colour variation is.

In the distance modulus sketched above, $\alpha$ and $\beta$ may be 
either global coefficients or depend on redshift, or depend on
the environment of the supernova (characterised for example 
by host galaxy colours). We also emulate a possible unnoticed evolution of 
supernovae intrinsic brightness (or a smooth redshift dependent distance bias) 
by allowing a redshift dependent intrinsic brightness:
\begin{equation}
\mathcal{M}(z) = \mathcal{M}_0 + e_M z
\label{eq:M_depend_z}
\end{equation}
where $\mathcal{M}_0$ and $e_M$ are parameters and $e_M$ is constrained
by a Gaussian prior (around a null fiducial value).
We will consider later several setups for $\alpha$, $\beta$ and
$\mathcal{M}$. We will use a fiducial value of $\sigma(e_M)=0.01$, 
and we sketch a scheme to constrain it from the survey in \S
\ref{sec:bias}. This modelling of the
``uncertainty floor''is significantly different from the assumptions
in \cite{Linder03}, where distance shifts in redshift bins are assumed
independent, and the uncertainty in a given redshift bin
depends on the highest redshift reached by the survey.
\subsubsection{Photometric calibration} 
Photometric calibration accuracy is naturally a major issue for supernova
surveys.  Supernova cosmology ``only'' requires relative calibration,
in the sense that the heart of the method consists in comparing the
flux of events across redshifts: cosmological results are
insensitive to the overall flux scale. 
Supernova fluxes are however to be measured in different bands: distant
supernova should be measured in redder observer bands than
nearby supernovae. Since supernovae fluxes are 
calibrated against standard stars, 
the calibration uncertainty  
arises in a first place from our limited knowledge 
of the fluxes of these standards, more precisely of the
ratio of their fluxes in different bands. A second contribution 
to flux calibration uncertainty arises 
from the measurement process itself, i.e. the systematic accuracy of 
supernova to standards ratio measurement.
We will examine later the influence of the uncertainty of
photometric calibration zero-points, where these uncertainties should
account for both sources. We typically assume that photometric zero points are
known to 1\% including the conversion to fluxes. This is conservative
in view of currently obtained precisions in the visible (see 
\citealt{Regnault09}, and \citealt{Guy10} for the total uncertainty).

\begin{table*}[!t]   
\begin{center}
\begin{tabular}{l|ccccc| c  c  c  D{.}{.}{-1} |}
{\bf label}     & {\boldmath $N_1$} & {\boldmath $N_2$} & {\boldmath $\sigma(zp)$}   & 
{\boldmath $\sigma_c$} & {\boldmath $\sigma(e_M)$ }  & {\boldmath $\sigma(w_a)$} & {\boldmath $z_p$} & {\boldmath $\sigma(w_p)$ } & \multicolumn{1}{c|}{{\boldmath $[\sigma(w_a)\sigma(w_p)]^{-1}$}} \\ 
\hline
A   &      0 &      0 &     0   &      0 &        0 &  0.23 &  0.41 &    0.020 &   213.9 \\ 
\hline
B   &     10 &      0 &     0   &      0 &        0 &  0.23 &  0.41 &    0.020 &   213.0 \\ 
\hline
C   &      0 &     10 &     0   &      0 &        0 &  0.23 &  0.41 &    0.020 &   213.9 \\ 
\hline
D   &      0 &      0 &  0.01   &      0 &        0 &  0.25 &  0.40 &    0.021 &   193.0 \\ 
\hline
E   &      0 &      0 &     0   &  0.025 &        0 &  0.25 &  0.41 &    0.020 &   198.1 \\ 
\hline
F   &      0 &      0 &     0   &      0 &     0.01 &  0.28 &  0.35 &    0.024 &   152.1 \\ 
\hline
G   &      0 &      0 &     0   &  0.025 &     0.01 &  0.29 &  0.36 &    0.024 &   141.7 \\ 
\hline
H   &     10 &      0 &     0   &  0.025 &     0.01 &  0.29 &  0.36 &    0.024 &   140.5 \\ 
\hline
I   &      0 &     10 &     0   &  0.025 &     0.01 &  0.29 &  0.36 &    0.024 &   141.7 \\ 
\hline
J   &     10 &     10 &     0   &  0.025 &     0.01 &  0.29 &  0.36 &    0.024 &   140.5 \\ 
\hline
K   &      0 &      0 &  0.01   &  0.025 &     0.01 &  0.32 &  0.37 &    0.027 &   117.0 \\ 
\hline
L   &     10 &      0 &  0.01   &  0.025 &     0.01 &  0.42 &  0.32 &    0.031 &    76.9 \\ 
\hline
M   &      0 &     10 &  0.01   &  0.025 &     0.01 &  0.32 &  0.37 &    0.027 &   116.9 \\ 
\hline
N   &     10 &     10 &  0.01   &  0.025 &        0 &  0.40 &  0.33 &    0.030 &    83.7 \\ 
\hline
Z   &     10 &     10 &  0.01   &  0.025 &     0.01 &  0.42 &  0.32 &    0.031 &    76.8 \\ 
\hline
\end{tabular}
\end{center}
\caption{Dark energy constraints when considering specific 
subsets of uncertainties. $N_1$ and $N_2$ refer to the number of 
parameters describing the supernova model (Eq. \ref{eq:polys}),
$\sigma(z_p)$ is the photometric calibration uncertainty for each band,
$\sigma_c$ is the scatter of light curve amplitude around the model, 
and $\sigma(e_M)$ is the prior applied to the evolution parameter $e_M$ defined in Eq. \ref{eq:M_depend_z}. 
$z_p$ denotes the (pivot) redshift which minimises 
the equation of state uncertainty, and this minimal uncertainty is labelled $\sigma(w_p))$.
The last column is the reciprocal of the area of the marginalised 
uncertainty ellipse ($\Delta \chi^2 = 1$) in the $(w_0,w_a)$ plane
(and also in the $(w_p,w_a)$ plane, see \citealt{DETF06}, p. 97).
\label{tab:acc-results}
}
\end{table*}

\subsection{Cosmological priors}
To forecast the measurement precision of supernova surveys
one has to complement the distance measurements by other 
constraints, because distances alone cannot efficiently separate
the various universe densities and the equation of state parameters. 
However, the parameter combination probed by distances to moderate redshifts
make them unique. We will complement the proposed
measurements with ``Planck priors''. For a geometrical probe like distances to SNe,
these Planck priors essentially consist in a single constraint
on the geometrical parameters ($\om$, $\Omega_{\rm DE}$, $w_0$, $w_a$).
Supernovae distances yield a second strong constraint in the same parameter
space, which is not enough to reduce the allowed region in 
the $w_0$,$w_a$ plane. Planck priors are more efficient to complement BAO
surveys for dark energy because CMB constrains a combination 
of geometrical parameters and sets the size of the BAO ``standard ruler''.
As an extra constraint for our supernova survey, we settle for the simplest one: flatness.
In using CMB priors for dark energy, one should pay attention to not
making use of dark energy information available through the ISW effect, 
because this information available on large scales
might not be reliably extracted. This is achieved by exactly enforcing 
the ``geometrical degeneracy'' (see e.g. \citealt{FOM-SWG09,Mukherjee08}).
Within this framework, using the full Fisher matrix or
a single one-dimensional geometrical constraint makes little difference.
Our ``Planck priors'' hence reduce to a measurement of the 
shift parameter R ($R \equiv \sqrt{\om H_0^2} r(z_{CMB})$) to 0.32\% 
relative accuracy (see \citealt{Mukherjee08}, Table 1), together with flatness.

\section{Results}
\label{sec:results}

We simulate a fiducial flat ${\rm \Lambda}$CDM universe 
with $\Omega_{\rm M}=0.27$. The equation of state is 
parametrised as $w(z) = w_0 + w_a\ z/(1+z)$. 
We assume that the scatter of the Hubble diagram for
perfect measurements exactly matching the average supernova model
(i.e without colour smearing) is 0.12 magnitudes. 
Current estimates of this quantity are below 0.10
\citep{Guy10}, but since it can only be obtained by subtraction of 
identified uncertainties that might have
been inadvertently inflated, we choose to stand on the safe side.
We assume 
that the colour smearing (see \S \ref{sec:photometric_precision_requirements})
causes peak SN magnitudes in a single band 
to scatter around the model by a quantity $\mathrm{\sigma_c=0.025}$
(see \citealt{Guy10}) unless otherwise specified.
$\sigma_c=0.01$ was assumed in \cite{Kim06}.

We restrict the rest frame central wavelength of the bands entering
the fit to [3800-7000]\AA, which leaves 3 to 4 bands per event.
Enlarging this range formally improves the performance but breaks the 
requirement that similar rest frame ranges are used to derive 
distances at all redshifts. 
In a real survey,
the whole information would of course be used, in particular
 to study the supernovae colours.

We will now study the survey performance regarding constraints of the equation
of state. Following \cite{DETF06},
we define the pivot redshift $z_p$ as the one where the EoS
uncertainty is minimal, and $w_p \equiv w(z_p)$. As performance
indicators, we report $\sigma(w_p)$, the uncertainty on of the EoS
evolution $\sigma(w_a)$, and the reciprocal of their product\footnote{
  This quantity is equal to $Det(Cov(w_0,w_a))^{-1/2}$, and scales as
  the figures of merit inspired from \cite{DETF06}.}, often used as a figure of merit (FoM).  The quantity
$\sigma(w_p)$ can be regarded as the ability of the project to
challenge the cosmological constant paradigm.

In Table \ref{tab:acc-results}, we turn various uncertainty sources on
and off labelled from A to Z. The setup A only considers photometry
Poisson noise and Hubble diagram scatter, and the baseline scenario Z
adds all other uncertainty sources discussed above. Some combinations
of uncertainty sources are provided in between, in order to identify
major uncertainty drivers. Lines B to F display the effect of one
source at a time : the single source that mostly degrades the performance is
the intrinsic brightness drift with redshift $\sigma(e_M)$, followed
by the photometric calibration ($z_p$).  In lines G to M, both intrinsic
drift with redshift ($\sigma(e_M)=0.01$) and colour smearing ($\sigma_c = 0.025$) are allowed, and we
add other sources. We note that adding the supernova model fit
alone (lines H,I,J compared to G) does not significantly degrade the performance and
that adding in the calibration uncertainty (K compared to G) has a 
sizable effect.  But
the most dramatic effect is the {\it combination} of supernova
model training and calibration uncertainty (lines H,K and L), because
the $N_1$ parameters of $p_1$ (Eq. \ref{eq:polys}) and the zero-points
play the same role, the first set in the supernova frame and the
latter in the observer frame.  This enlights the role of calibration
uncertainty through the light curve fitter training, and is supported
by the findings of current ground-based surveys \citep{Guy10}. Turning
on or off the colour law fitting ($N_2$, comparing lines G and I, 
or  L and Z) has essentially no impact on
the performance : there is indeed no benefit to rely on an assumption
(such as a Cardelli law)  for this part of the model.

One might be surprised that calibration uncertainties do not ruin the
proposed survey since current ground-based supernova surveys face
calibration-induced uncertainties comparable to statistics with ``only'' 
a few hundred events \citep{Guy10, Conley10}. There are two key differences:
first, these survey have to face cross-calibration issues with the
nearby sample; second, the observed rest frame region varies with
redshift, and the comparison of events across redshifts then heavily
relies on the supernova model and inherits its uncertainties. 
This illustrates our requirement that
all events be observed in the same rest frame range.

Comparing colour distributions along redshift 
constitutes a sensitive handle on astrophysics conditions 
evolving with time.
This test is independent of the supernova model training,
in the sense that the training does not aim at matching 
these distributions. The photometric calibration accuracy
directly limits the comparison of colours at different redshifts:
with zero-points defined at $0.01$, colours can only
be compared to $\sim 0.015$. Given the natural colour spread of 
$\sim 0.1$, this is a test at $\sigma/7$ level, which does not 
benefit from more than $\sim 50$ events in a redshift slice.
For supernovae, the sensitivity of evolution tests constitutes another 
strong incentive to improve the calibration precision to a few per mil
level.

\subsection{Variations of performance when altering parameters 
of the baseline.}
\label{sec:variations}

\begin{table}[ht]
\begin{center}
\begin{tabular}{l|cc|}

{\bf Alteration} &  {\boldmath $\sigma(w_p)$} & {\boldmath $[\sigma(w_a)\sigma(w_p)]^{-1}$} \\  
\hline
\textbf {none}      &    \textbf{0.031} &    \textbf{76.8 }\\ 
\hline
$N_1 = 15$      &    0.031 &    75.4 \\ 
\hline
$N_2 = 15$      &    0.031 &    76.8 \\ 
\hline
$\sigma_{int} = 0.10$      &    0.029 &    86.7 \\ 
\hline
$ \sigma(e_M) = 0.02 $      &    0.033 &    64.6 \\ 
\hline
$ \sigma(zp) = 0.005 $      &    0.028 &    99.3 \\ 
\hline
$ \sigma_c = 0.015 $      &    0.029 &    86.3 \\ 
\hline
half statistics      &    0.037 &    50.6 \\ 
\hline
double statistics  &    0.025 &   118.1 \\ 
\hline
only wide      &    0.044 &    39.2 \\ 
\hline
only deep      &    0.046 &    31.9 \\ 
\hline
wide $ \times$ 0.82, deep $ \times$ 1.25      &    0.031 &    74.7 \\ 
\hline
wide $ \times$ 1.18, deep $ \times$ 0.75      &    0.031 &    78.0 \\ 
\hline
\end{tabular}
\end{center}
\caption{Effect of altering survey parameters of the baseline
(defined as line Z of Table \ref{tab:acc-results}).
\label{tab:delta-baseline}}
\end{table}

\begin{figure}[h]
\centering
\includegraphics[width=\linewidth]{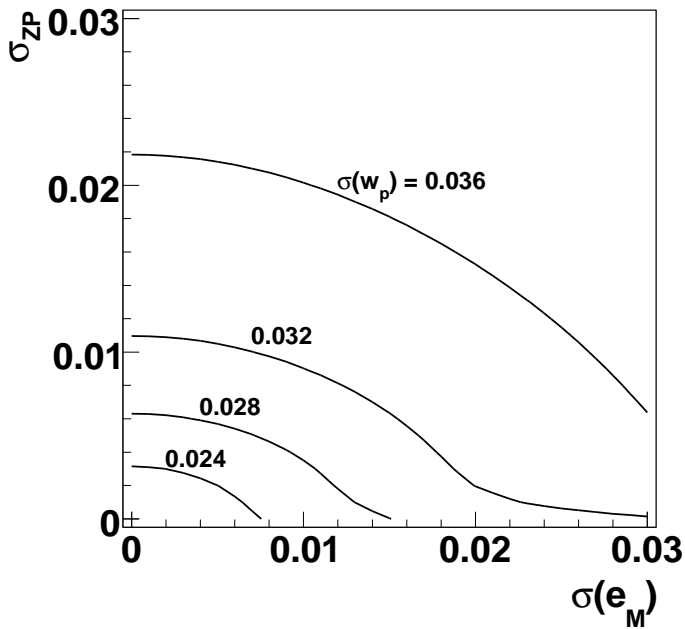}
\caption{Contour levels of $\sigma(w_p)$ as a function of calibration 
uncertainty $\sigma_{ZP}$ (equal for all filters), and the
distance evolution uncertainty $\sigma(e_M)$ (defined in Eq. \ref{eq:M_depend_z}).
\label{fig:wp_contours}
}
\end{figure}

Table \ref{tab:delta-baseline} illustrates the effect of parameters
that determine the cosmological performance, and varies them one at a
time. We can check that increasing the spectral resolution of the
supernova model ($N_1$ and $N_2$) has almost no effect. If an
intrinsic resolution $\sigma_{int}$ better than 0.10 is confirmed,
this yields a $\sim$ 10 \% improvement of the FoM.  
Doubling the evolution systematics
$\sigma(e_M)$ degrades it by 16~\%. Improving the zero-points
accuracy by a factor of 2 improves it by 30~\%. The joint effect of
calibration uncertainty and distance biases on $\sigma(w_p)$ is
displayed in Fig. \ref{fig:wp_contours}.  One might also note 
that reducing statistics significantly alters the performance.
Conversely, doubling the statistics improves the FoM by 50~\%. 
We should probably stress here that the wide survey should
not be regarded as doable from the ground, because supernova
distances at $z\sim 1$ make use of y and J bands.
We finally vary the wide and deep survey allocations within a 
constant overall observing time, and note that this is not a key parameter.
Reducing the deep survey marginally improves the FoM but at the expense of 
reducing the high redshift statistics, which is the most precious 
to tackle evolution issues. One might also regard the rest frame I-band 
Hubble diagram as an extension of the wide that would improve the constraints.
If, following the SNAP approach, one includes 1000 nearby supernovae 
at $z=0.05$ (measured from the ground), 
assuming that cross-calibration uncertainties are not worse than assumed 
here, and ignoring potential bias issues, the FoM reaches 110.

One should seriously consider the possibility that there are
subclasses of SNe~Ia depending on environment as suggested by
\cite{Mannucci06}. \cite{Sullivan06,Sullivan10} propose some observational
evidences for different average properties of SNe occurring in passive
and star-forming galaxies. It is not yet clear if these different
environments produce different supernovae, or if these different
environment sample differently the same parent population. We will
consider however here the most dramatic case, where two environments
produce two different event species described by different parameter
sets, namely different $\alpha$, $\beta$, $\mathcal{M}$ (see equation
\ref{eq:m-alpha-beta}) and different supernova models (different $p_1$
and $p_2$ functions, see equation \ref{eq:polys}). We consider that
the host galaxy colours allow one to assign each event in one
category.  In the case of the admixture not evolving with redshift and
categories having the same photometric quality, 
the variance of the cosmological
estimators is mathematically the same as for a single species scenario,
as shown in Appendix \ref{app:several_event_species}.  For a more
realistic scenario, we varied the admixture with redshift from 30/70\%
at $z=0$ to 70/30\% at $z=1.5$, and variances of the cosmological
parameters do not increase by more than 1\%.

So, if it turns out that SNe~Ia consist in an admixture of species
that can be tagged using host galaxy colours or properties of the
supernova light curve, the proposed strategy retains its
performance. Note that even if events are assigned the wrong category,
the cosmological parameters remain unbiased as long as the fraction of
interlopers does not evolve with redshift.  We also considered the
possible evolution with redshift of $\alpha$ and $\beta$ parameters
(defined in equation \ref{eq:m-alpha-beta}) by fitting those in
redshift bins of 0.1, and variances are changed at the percent level. Even
considering simultaneously event sub-classes and redshift bins
results in minute degradation of performances.

Finally, we replaced the evolution of $\mathcal{M}(z)$ defined in Eq. \ref{eq:M_depend_z} by:
$$
\mathcal{M}(z) = \mathcal{M}_0 + e_M \frac{ [ \partial Log d_L/\partial w_0 ] (z)}
{[ \partial Log d_L/\partial w_0 ] (z=1)}
$$
where the denominator has been chosen so that $\sigma(e_M)$ represents
the uncertainty at $z=1$ as in Eq. \ref{eq:M_depend_z}. This functional dependence makes $e_M$ degenerate with $w_0$. The  choice
of a systematic mimicking the dependence on $w_0$ was inspired by
\cite{Amara08} where, in the weak shear framework,  it is regarded 
as a worst case. However, in our case, for the same strength of the prior, 
we find that the FoM are larger (or equal) than with the 
linear model of Eq. \ref{eq:M_depend_z}.

\subsection{Constraining a redshift dependent bias of distances}
\label{sec:bias}
As discussed in the previous paragraph, cosmological constraints 
from supernovae depend on our ability to constrain the systematic drift
of distances with redshift. We investigate in this section a few handles
that might be used to bound the effect or correct for it.

The proposed survey provides accurate colours of host galaxies
extending to the observer NIR. This enables one to estimate host galaxy
properties by comparing measurements to synthetic galaxy spectra,
and analyse if derived distances (or other supernovae properties) depend on 
the host galaxy properties. This is the strategy followed
 in \cite{Sullivan10}, where evidence for a dependence of distances with host 
galaxy stellar mass is presented, and found harmless for cosmology.
With the supernova statistics we are considering here, 
this approach becomes even stronger because it can be applied within
a modest redshift range where photometric calibration issues affect
all supernovae and host galaxies in the same way.

The metallicity of exploding white dwarves is expected to increase
(at least on average) with cosmic time. Metallicity certainly
influences the amount of $^{56}\mathrm{Ni}$ synthetised in the 
explosion \citep{Timmes03}, but we have no hint yet that standard distance 
estimators do not correct for this evolution. One might even
argue that the range of environments found at, e.g., small redshifts
efficiently ``trains'' distance estimators so that they remain unbiased
as redshift varies. 

   However, in order to bound a possible redshift dependent distance 
bias we propose to use the near UV flux variations that
explosion models correlate to metallicity variations
\citep{Hoeflich98,Lentz00,Sauer08}. If the models 
do not necessarily agree on the size of the effects, they
define the 2500-4000 \AA\  spectral region as sensitive to
admixtures in the progenitor material of other elements than 
Carbon and Oxygen (the ones of the baseline scenario). 
SNe~Ia exhibit a colour diversity, but different colours appear to
be tightly connected (see e.g \citealt{Astier06,Conley08,Folatelli10}).  
Defining $U^*$ as the 2500-3200\AA\ \ region, the rest
frame combination $(U^*-B)-5.4(B-V)$ is, from observations, the smallest scatter
combination of the $(U^*,B,V)$ triplet \citep{Guy10} (up to a multiplicative constant). Using synthetic
data from \cite{Lentz00}, we can check that this three-band
combination is both sensitive to metallicity, and nicely correlates to
distance biases due to evolving metallicities. Quantitatively,
when varying metallicity, the
distance shift $\delta\mu$ varies as 
$0.1[( U^*-B)-5.4(B-V)]$, which
allows one to constrain $\delta\mu$ at the 0.01 level in the presence of
calibration uncertainties of 0.01 per band. The $U^*$ band is
observable in the proposed survey beyond $z=0.6$, and 
lies below the supernova model spectral region, and 
is then ignored in the training.

Beyond the decline rate paradigm, the early phases of
light curve are expected to encode metallicity (see \citealt{Hoeflich98}).
Given the envisaged statistics, minute departures from the average
light curve shape can be detected and correlated with other
observables. One key quality of light curve shape indicators is that
they remain unaffected by calibration issues.

\section{Discussion}
\label{sec:discussion}
\subsection{Comparison with weak shear performance}
The ability of space based measurements of the weak shear 
to constrain the cosmological model, and in particular its dark energy sector 
was studied in detail in \cite{Amara08} (and references therein). 
This work supports the strategy developed for the EUCLID project 
\citep{EuclidYB}, and identifies the measurement of the shear
(intimately related to the measurement of second moments of galaxies)
as the key systematic. The shear measurement uncertainty translates to
a systematic uncertainty floor of the shear angular power spectrum $C^{sys}_\ell$.
Current shear measurement techniques 
\citep{Great08Results}
achieve a systematic uncertainty that would limit $\sigma(w_p)$ to about
0.05 (Fig 11 of \cite{Amara08} interpolated 
for $C^{sys}_\ell \simeq 2\ 10^{-6}$ from the best algorithm in 
Table 5 of \citealt{Great08Results}). One should note 
that this performance is obtained on simulations
where in particular, the PSF of the imaging system 
used to measure the shear is perfectly known. The EUCLID project
requires that $C^{sys}_\ell < 10^{-7}$ be reached \citep{EuclidYB}, leading to  $\sigma(w_p) \simeq 0.02$.

Using the current performance of ground-based distance measurements 
to supernovae, 
we conservatively derive an EoS constrain 
$\sigma(w_p)\simeq 0.03$ from a space-based
survey. This is  in a position to really complement EoS constraints
from shear correlations.
These two approaches are not only complementary because their redshift
dependent biases are unrelated. On the one hand, distance
tests are independent of growth rate, and have to be complemented
(by e.g. Planck priors) in order to constrain the EoS,
and on the other hand, shear correlation tomography can autonomously
constrain the EoS by assuming a given relation (from e.g. General Relativity) 
between distances and growth rate. 

\subsection{Summary}
We have proposed a two-cone supernova survey conducted from space with
a modified EUCLID setup : we assumed that both the visible and the NIR
imagers are equipped with swappable filters. We find that it is
possible to accurately measure more than $10^4$ supernovae at
$0.15<z<1.55$ in 18 months of survey. The photometric accuracy is
tailored to match the measured intrinsic variability of colour
relations of supernovae at the highest redshift of the survey.  All
events are measured in the 7 instrumental bands, and the BVR rest
frame bands are covered at all redshifts. Our analysis of supernova
distances relies on
conservative assumptions and the current know how. It integrates many
nuisance effects, such as the light curve fitter noise together with
the impact of a conservative photometric calibration uncertainty both
directly on cosmology and through the light curve fitter training. Our
approach ensures that the interplay of identified uncertainties is
properly accounted for. The proposed observing strategy also collects
the data to build a rest frame I-band Hubble diagram to $z\simeq 0.9$,
with $\sim$ 7000 events.

Our results are encouraging in the sense that including these
realistic nuisance effects, competitive constraints of the dark energy
equation of state can be obtained, when using a simple geometrical
Planck prior: within a two-parameter dark energy model, the EoS can be constrained
to $\sim$0.03 at $z\simeq 0.3$.

\begin{acknowledgements}
We are grateful to Peter Nugent for providing us with the
simulated spectra discussed in \cite{Lentz00}, and to Andrew Howell
for fruitful exchanges.
\end{acknowledgements}

\appendix
\section{Simultaneous versus successive least squares fits}
\label{app:succ-simul}
We compare here two approaches for the least-squares fits in the
supernova framework. In the traditional approach one first fits 
separately the event light curves to extract the event parameters, and
then fits the cosmology to distances constructed from these event
parameters. In the less conventional approach we present here, light
curves and cosmology are fitted simultaneously, by minimising the sum:
\begin{equation}
\chi^2(p_c, \theta)  = \chi^2_c + \sum_{i=1}^{N_{sn}} \chi^2_{LC,i}
\label{eq:chi2g}
\end{equation}
with
\begin{align}
\chi^2_c & =  \sum_{i=1}^{N_{sn}}  \frac{(\mu(p_c,z_i) - h_c^T\theta_i -\mathcal{M})^2}{\sigma_{int}^2} \label{eq:chi2c}\\
\chi^2_{LC,i} & =  \sum_{k=1,N_i} \frac{(f_{ik} - \phi(\theta_i, t_{ik}))^2}{\sigma_{ik}^2} \label{eq:chi2lci}
\end{align}
where, $i$ indexes SN events,  
$k = 1..N_i$ indexes the measurements $f_{ik}$ of a an event, 
$\phi$ is the supernova model, $\theta_i$ are the parameters of event $i$
($\theta$ denotes the ensemble of event parameters),
$p_c$ are the cosmological parameters, $\mu(p_c,z)$ is the distance 
modulus at redshift z, $h_c^T\theta_i$ is the measured distance modulus
of event $i$ (a linear
combination of event parameters), and $\mathcal{M}$ is a combination
of the intrinsic magnitude of a supernova and $H_0$.
$\sigma_{int}$ is the intrinsic dispersion of a supernova (defined as the ``observed'' Hubble diagram scatter for ideal measurements), and $\sigma_{ik}$
refers to measurement uncertainties. The first $\chi^2$ term fits cosmology,
the second fits light curves, and both terms are related through 
$\theta_i$ parameters.

If one minimises separately the light curve parts with respect to $\theta_i$,
$\chi^2_{LC,i}$ may be approximately re-written:
\begin{equation}
\chi^2_{LC,i} = (\theta_i-\theta_i')^TW'_i(\theta_i-\theta'_i)
\end{equation}
where $\theta_i'$ minimises $\chi^2_{LC,i}$ and $W'_i$ is its second
derivative matrix w.r.t $\theta_i$. The approximation would be exact if the
supernova model were linear with respect to its parameters. The
approximation holds for error propagation at first order, and will be
used only for this purpose. In order to compare the simultaneous fit
of $p_c$ and all $\theta_i$ with the two-step process, we will
compute $\chi^2(p_c, \hat \theta)$ where $\hat\theta$ denotes the
$\theta$ value that minimises the global $\chi^2$ (\ref{eq:chi2g}) for a given value of
$p_c$. A tedious calculation yields:
\begin{equation}
\label{eq:chi2c2}
\chi^2(p_c) = \sum_{i=1}^{N_{sn}}  \frac{(\mu(p_c,z_i) - h_c^T\theta'^i -\mathcal{M})^2}{h^T W'_i  h+\sigma_{int}^2} 
\end{equation}
which corresponds to the standard cosmology fit that one would perform
after a separate fit of all events to yield $\theta'_i$ and $W'_i$. So,
fitting all events and the cosmological parameters simultaneously
(using expressions \ref{eq:chi2c} and \ref{eq:chi2lci}) yields exactly
the same cosmological estimators as the ``traditional'' two-step method.
Note that the expressions \ref{eq:chi2c} and \ref{eq:chi2c2} have
in particular different denominators, indicating that the 
simultaneous fit is a convenient way of propagating uncertainties 
of other analysis steps. This result is very general, and
should be no surprise to Kalman filter practitioners.

\section{The local-global technique}
\label{app:local-global}
For a supernova sample of several thousand events, the number of 
parameters in expression \ref{eq:chi2g} becomes uncomfortably large
(even for simulations), at least when it comes to inverting the 
$\chi^2$ weight matrix. However, since no $\chi^2$ term involves
the parameters of two different events, this weight matrix is sparse.
We write it block-wise:
\begin{equation}
W \equiv  \begin{pmatrix} 
W_p    & A_1 & \cdots & A_i &\cdots  & A_N \\ 
A_1^T  & W_1 &        & 0   &        & 0   \\
\vdots &     & \ddots & 0   &        & 0   \\
A_i^T  & 0   &        & W_i &        & 0   \\
\vdots &     &        &     & \ddots & 0   \\
A_N^T  & 0   &  0     &  0  &   0    & W_N 
\end{pmatrix}
\nonumber
\end{equation}
The first row and column refer to $p_c$ and the following ones to 
the successive $\theta_i$. It is easy to check that the corresponding
blocks of the inverse matrix read:
\begin{align}
C_p &= (W_p-\sum_{i=1}^N A_iW_i^{-1}A_i^T)^{-1} \label{eq:C_p} \\
C_{p,i} & =  -C_p A_i W_i^{-1} \nonumber\\
C_{i,j} & =  \delta_{ij}W_i^{-1} + W_i^{-1}A_i^TC_pA_jW_j^{-1}  \nonumber
\end{align}
The solution of the linear system:
\begin{equation}
W \begin{pmatrix} p \\ \theta_1 \\ \vdots \\ \theta_N  \end{pmatrix} =
  \begin{pmatrix} B_p\\ B_1 \\ \vdots \\ B_N  \end{pmatrix} 
\nonumber
\end{equation}
can be written in the efficient form:
\begin{align}
p &=  C_p(B_p-\sum_{i=1}^N A_i W_i^{-1}B_i) \nonumber \\
\theta_i & =  W_i^{-1}(B_i-A_i^T p)  \nonumber
\end{align}
which only requires the $C_p$ block from the inverse.
Note that the computing time
scales as the number of events N, rather than $N^3$ if some standard
factorisation of $W$ were used. In our case, the $C_p$ matrix is obtained 
in matter of seconds with O(10000) supernova events. 

In order to fit in this scheme, least squares problems should exhibit
``local'' ($\theta_i$) and ``global'' ($p$) parameters. By definition,
the local parameter sets $\theta_i$ are not connected to each other through any
$\chi^2$ term, and the single global parameter set $p$ gathers all the
remaining ones. In the cases discussed in this work, the split is
fairly obvious: local parameters refer to supernova event parameters
and the global set $p$ contains everything else (cosmology,
zero-points, ...). This block solution scheme of our least squares minimisation was proposed for a totally different purpose in  
\cite{Regnault09}.

\section{Several event species}
\label{app:several_event_species}
We prove here why splitting the supernova sample into several event
species does not degrade the cosmological parameters, if these species
have the same redshift distribution and photometric accuracy. We
formally split the global parameters (all parameters but the supernova
events ones), into the common ones (labelled by $c$) which typically gather
cosmology and zero-points, and the parameters specific to each species
(labelled by $1$ and $2$). The later gather the supernova model and the
distance estimator parameters. The global parameter space addresses
$(g_c,g_1,g_2)$, and we wish to compute the covariance matrix of $g_c$
(which contains the cosmology) when fitting both samples simultaneously. 
We add the inverses of both samples covariance
matrices, expressed in the proper subspaces:
\begin{align}
W_{1+2} = W_1+W_2 & = 
\begin{pmatrix}
W_{c1} & X_1 & 0 \\
X_1^T & W_1 & 0 \\
0 & 0 & 0  
\end{pmatrix} +
\begin{pmatrix}
W_{c2} & 0 & X_2 \\
0   & 0 & 0 \\
X_2^T & 0 & W_2 
\end{pmatrix} \nonumber \\ 
    & = 
\begin{pmatrix}
W_{c1}+W_{c2} & X_1 & X_2 \\
X_1^T & W_1 & 0 \\
X_2^T & 0 & W_2
\end{pmatrix} \nonumber
\end{align}
Using Eq. \ref{eq:C_p}, the covariance matrix of the common parameters reads:
$$
Cov(g_c) = \left [ (W_{c1}+W_{c2})-X_1^TW_1^{-1}X_1-X_2^TW_2^{-1}X_2 \right ]^{-1}
$$
If both species have the same properties (redshift distribution,
photometric quality), we have $W_1 = k W_2$  and $X_1 = k X_2$,
where k stands for  the population ratio. So,
\begin{multline*}
Cov(g_c) = \left [ (W_{c1}  +W_{c2})- \right . \\ 
	 \left (X_1+X_2)^T (W_1+W_2)^{-1}(X_1+X_2)   \right ] ^{-1}
\end{multline*}
This expression describes as well
all events belonging to a single species, which completes the proof.
As expected, the argument breaks down if there are not enough 
events to constrain $g_1$ or $g_2$, i.e. if $W_1$ or $W_2$ is singular.
One may also note that the argument applies as well to any number
of event species.

\bibliographystyle{aa}
\bibliography{biblio}

\end{document}